# Energy Consumption Forecasting for Smart Meters

Anshul Bansal, Susheel Kaushik Rompikuntla, Jaganadh Gopinadhan [1], Amanpreet Kaur, Zahoor Ahamed Kazi

Cognizant Technology Solutions
Bangalore, India
{Anshul.Bansal, SusheelKaushik.Rompikuntla,
Jaganadh.Gopinadhan,Amanpreet.Kaur2,
ZAHOORAHMED.KAZI}@cognizant.com

## ABSTRACT

*Earth, water, air, food, shelter and energy are essential factors required for human being to survive on the planet. Among this energy plays a key role in our day to day living including giving lighting, cooling and heating of shelter, preparation of food. Due to this interdependency, energy, specifically electricity, production and distribution became a high tech industry. Unlike other industries, the key differentiator of electricity industry is the product itself. It can be produced but cannot be stored for future; production and consumption happen almost in near real-time. This particular peculiarity of the industry is the key driver for Machine Learning and Data Science based innovations in this industry. There is always a gap between the demand and supply in the electricity market across the globe. To fill the gap and improve the service efficiency through providing necessary supply to the market, commercial as well as federal electricity companies employ forecasting techniques to predict the future demand and try to meet the demand and provide curtailment guidelines to optimise the electricity consumption/demand. In this paper the authors examine the application of Machine Learning algorithms, specifically Boosted Decision Tree Regression, to the modelling and forecasting of energy consumption for smart meters. The data used for this exercise is obtained from DECC data website. Along with this data, the methodology has been tested in Smart Meter data obtained from EMA Singapore. This paper focuses on feature engineering for time series forecasting using regression algorithms and deriving a methodology to create personalised electricity plans offers for household users based on usage history.*

## INTRODUCTION

Earth, water, air, food, shelter and energy are essential factors required for the human beings to survive on the planet. Among this, energy plays a key role for our day to day living, including giving lighting, cooling and heating of shelter, preparation of food and other day to day activities. Due to this interdependency, energy, specifically electricity, production and distribution became a high tech industry. Unlike other industries the key differentiator of electricity industry is the product itself. It can be produced but cannot be stored for future; production and consumption happen almost in near real-time. This particular peculiarity of the industry is the key driver for Machine Learning and Data Science based innovations in this industry. There is always a gap between the demand and supply in electricity market across the globe. To fill the gap and improve the service efficiency through providing necessary supply to the market; commercial as well as federal electricity companies employ forecasting techniques to predict the future demand, try to meet the demand and provide curtailment guidelines to optimise the electricity consumption.

---

[1] Corresponding Author
Jaganadh Gopinadhan
Jaganadh.Gopinadhan@cognizant.com

With the ever increasing and fluctuating drivers towards the energy consumption patterns it is a quite challenging task for Machine Learning researchers to build a single stop modelling solution for electricity consumption forecasting for individual users. The consumption patterns of electricity depend on the user type such as individual, office or industry. The introduction of smart meters to measure the consumption in 30 minutes' granularity have provided the greatest power to Data Scientists to study the usage patterns and device the best solution for industry including forecasting, personalized electricity plans and curtailment suggestions. Typically, the consumption is depending upon socio-economic and climatic conditions of a region under consideration for the forecasting exercise. Even though the relation with this drivers and consumption looks straight forward, there are many interesting patterns available in the data which is influenced by any one or more of the above factors. To understand this better, one of the key patterns in household electricity consumption related to holidays is 'day shoulder effect'. The day shoulder effect is the influence of a special occasion/holiday in preceding days and or following days. The obvious example is Christmas and New Year celebrations time. The house hold users tend to illuminate more lights during the season; even days before the actual holiday and these days' consumption pattern varies on parts of the days such as evening and nights. There can be some patterns visible in the data based on the professional profile of the residents in a building. Typically, a residence with students, youngsters will be having comparatively high electricity consumption during the night; at the same time a group of working professional's residence will be having high energy consumption during the evening and early morning. These behavioral, social and economic aspects of human life impact the consumption, hence capturing these features directly or indirectly helps the Data Scientist to build an effective model for both consumption and load forecasting scenarios in electricity industry.

The current exercise focuses on investigating the use of regression techniques combined with effective feature engineering techniques to forecast the individual user's energy consumption. The data used for this exercise is obtained from the Europe's Public Data, the DECC live data[1]. Apart from this, some private data from EMA Singapore was collected as part of the study. The results from the same is not included in the paper. As of now the team is conducting more experiments with the UK Government public data[2] to expand this methodology. The algorithm used for this exercise is Boosted Decision Tree Regression implemented in Microsoft AzureML[3] platform. The following sections of the paper discuss the current state of the electricity demand/load forecasting techniques and approach on modelling. The comprehensive feature engineering and suggestions on feature engineering and operationalizing the forecast for electricity plan personalization is also discussed in this paper.

---

[1] https://publicdata.eu/dataset/decc-live-energy-data  
[2]https://data.gov.uk/dataset/energy-consumption-for-selected-bristol-buildings-from-smart-meters-by-half-hour  
[3] https://studio.azureml.net/ , https://msdn.microsoft.com/en-us/library/azure/dn905801.aspx

# 1 LOAD FORECASTING APPROCHES

## 1.1 Load Forecasting

Load Forecasting techniques are one of the widely studied aspects using various time series techniques and regression algorithms including ARMA and Neural Networks. Most of the studies in this direction use statistical techniques or Artificial Intelligence (AI) algorithms such as regression, neural networks, fuzzy logic and expert systems. Apart from this there are some naïve techniques such as similar day approach i.e. copying the same day from previous year. Most of the statistical approaches leverage a mathematical model, which can represent a load as function. The most widely used such methods are additive and multiplicative models. Chen, Canizares and Singh [1] presented an additive model in their paper. A rule based short term forecasting technique was proposed by Rahman [2]. An ARIMA Transfer Function based approach is discussed in the paper by Cho et al [3]. Neural Network and its derivatives are being widely studied in electric load forecasting. A short-term forecasting method is presented on the basis of RBF neural network and fuzzy reasoning is discussed in the paper by Yun Lu and Yinuo Huang [4]. Optimization techniques for short term load forecasting with Fractal Theory is studied by Yongli et al [5]. An ANN based similar day approach is being discussed by Chong Yin et al in their paper [6]. ARIMA based short-term and long-term electricity load forecasting of Shanghai Urban Area is discussed in the paper by Xie et al [7]. Application of Fuzzy Support Vector Machine Regression with temperature information for load is discussed in the paper by Changyin Sun et al [8]. A genetic and swarm optimization algorithm for energy demand forecasting using the economic factors is studied by Hesam Nazari et al [9]. An interesting study on using Grey Forecasting methodologies for load forecasting is discussed by Yan Yan et al [10]. Yang et al discussed a combination forecasting method with neural network [11]. A Case Based Reasoning Approach System for load forecasting is discussed by Raul et al in their paper [12]. Correlation Dimension Estimation to estimate the model order of electrical load data with neural network is being studied by Francesco et al [13]. The paper majorly discuss the model order and designs appropriate neural network based model for forecasting. An improved Practical Swarm Optimizer and Least Square Support Vector Machine based short term load forecasting is studied by Qianhui Gong et al [14]. A detailed discussion on common load forecasting techniques is being discussed in the paper "Load Forecasting" by Joe H et al [15] and Hoessian et al [27].

## 1.2 Feature Selection for Time Series Forecasting

Efficient feature selection and engineering is key for any forecasting or modelling exercise. Most of the papers discussed in the previous section use different features for the modelling exercise. The raw data for energy forecasting have information such as date, time and load/consumption. In addition to this, weather data, calendar information etc. are used along with the raw data to make the prediction much reliable. More than the directly available features, lots of engineered features are included in the exercise, which shall be discussed in detail in forthcoming sections.

Feature selection in time-series data is one of the emerging and challenging area in Machine Learning. In the Big Data and IoT revolution era real-time time-series data is captured and various statistical analysis and forecasting are applied. Feature selection with reference to electric load forecasting is discussed by Mashud et al in the paper [16]. The paper discussed measures such as Prediction Interval Quality, LUBE and improved LUBE (LUBEX), Mutual Information and Correlation Based Feature Selection. A generic filter and wrapper based approach for neural network based time-series forecasting is studied by Sven F et al [17]. Automatic feature generation using Grammar Based techniques are discussed in the seminal book by De Silva et al [18].

## 2 DATA

Half hourly electricity consumption data from the DECC headquarters building at 3-8 Whitehall Place UK for the primary experiment. The data is available at European data initiate web portal [19]. Apart from this an experiment with Smart Meter data from EMA Singapore was conducted to understand the electricity usage patterns. Results from the same is not discussed in the paper in detail as the data is not open for public sharing. As a follow-up study to this experiment, a set of supplementary exercises with application of deep learning methodologies in load/consumption forecasting with the UK Government open data [20]. The supplementary data used for incorporating additional features in the experiment included a generic holiday list gathered from various internet resources.

The original data consists the following attributes:

*Table 1: Attributes in the Data*

| Attribute Name | Description | Remarks |
| --- | --- | --- |
| Site Name | Name of the building | Not used in the model |
| Utility | For which utility the measurement is taken. | Not used in the model |
| Unit | Measurement unit | kWh |
| Consumption from 00:00 to 23.00 | Every 30 mins consumption | Used in the model |
| Total | Total energy consumption for a day. | Not used in the model |
| Date | Date which the measurement is taken | Used in the model |

The holiday gathered from various sources have the following attributes: Date, is holiday (Boolean value) and name of the holiday. Both the data were joined to get the first set of data for feature engineering.

### 2.1 Feature Engineering

Feature engineering is at the heart of any data science project in which the Data Scientists create features that make machine learning algorithms work. The typical steps involved in feature engineering and feature extraction include,

  a) Brainstorming features: during this step a subject matter expert in the industry will be working with the Data Science team to provide deep insight about the business process.
  b) Feature designing: during this phase new features will be constructed from the existing data or combining additional data from various sources applicable.
  c) Feature Extraction: in this phase the Data Scientist validates the predictive power of new features as well as existing features. There are many techniques applied to validate the feature importance such as correlation analysis, ensemble and tree based model based feature importance, Recursive Feature Elimination (RFE), and Recursive Feature elimination with Cross validation (RFECV) etc.
d) Feature Transformation/Derivation: during the validation with a baseline model some of the feature may require transformation. These transformations include log transformation, Standard Scaling (SS) and Min Max Scaling (MMS).

After literature survey and consultation with subject matter expert, a set of most desirable features for electricity load/consumption forecasting were listed.

a) Past consumption pattern: electrical consumption pattern cannot change abruptly until unless some major changes happen at the place. So past consumption pattern carries information for future consumption pattern.
b) Calendar: Season, month, day of week, weekend
c) Events around the location: Football or cricket match may increase television watching and thus increase the consumption. Similarly, events like Christmas change the consumption of electricity.
d) Demography: The population of building, age group, office or house hold, gender etc. such information can also affect the consumption pattern.
e) Geography: Locality, temperature, rain, weather, etc. If temperature is high, people will use more electrical appliance and similarly when temperature is low.

In feature designing step, the availability of data to build features to encapture above mentioned was checked and then relevant features were designed. After designing the logical features, the next step is to extract the most important and independent features out of it. For example, if we look at the history based feature there can be lot of features that we can design like consumption on day-1, day-2, day-3, week-1, week-2, week-3 and so on. In feature extraction step we performed autocorrelation and found the most correlated variables among all. The feature set was a mix of categorical and numerical features. The numerical features were in varying range. So log transform is performed in order to scale the data.

The extracted and pre-existing features were classified into three categories and form the data used for this experiment.

a) Calendar based features: day, month, week day or weekend, time part such as morning evening and night.
b) Event based features: holiday
c) Load history based features: Current half hour minus one to ten half hour consumption and minimum and maximum consumption of the previous day in this feature set. Along with this the current day minus one-week consumption is also considered.

After developing a feature the validity of its predicting power is tested by calculating its statistical properties. E.g., the consumption of electricity for a household depends on the history of consumption. So several consumption history based features were developed but when tested only few of them had the predictive capacity, so only few features were considered in final model, such as current half hour consumption with previous three half hour consumption. Features such as previous week consumption were excluded from the data, due to lack of predictive power.

## 2.2 Handling Missing Values

Pre-processing in machine learning never follow a linear path. The data is formatted by taking the transpose of the day load and then all day loads are merged with proper period index. 25% of the load data had missing values. A linear fitting approach has been adopted to fill these missing entries. The missing values can be because of a power cut or because of the error in recording. Brsitol buidling area do not have regular power cut power, cut so the 25% missing values were consired as loast data. For short duration missing values, assuming the consuption will not be higly turbulent, a linear extrapolation has been done . This extrapoltion is done with Run Length Encoding (RLE). If the duration is longer than couple of hours, then average of week-1 and day-1 was used to fill the missing value.

## 3 MODELLING OF ENERGY CONSUMPTION

Before starting the formal exercise, some minor experiments with Auto Regressive Integrated Moving Average (ARIMA) and Exponential Moving Average (EMA) were conducted. These techniques were not able to accommodate the categorical features included in the data such as month etc. Considering the categorical feature in the data-set an ensemble method Boosted Decision Tree Regression (BDTR). The selected implementation is AzureML based one which leverages the LambdaMART [21].

### 3.1 Boosted Decision Tree Regression (BDTR)

Regression Trees are a set of supervised learning methods, which try to address multiple regression problems. This model provides a tree based approximation $\hat{f}$, of an unknown function $Y = f(x) + \varepsilon$ with $Y \in R$ and $\approx N(0, \sigma^2)$ based on a given training data $D = \{\langle x_{i,1,\dots} x_{i,p}, y_i \rangle\}_{i=1}^{n}$ . The trees obtained from the model will consist of a hierarchy of logical tests on the value of any of the p predictor variables. The leaves a.k.a terminal nodes of the trees will contain the numeric prediction of the model for the target variable. A seminal work in this area is being produced by Brieman et al [28]. The key characteristics of this model is automatic variable selection, computational efficiency, handling unknown attributes and categorical data and interpretability of the model. Despite of all of these advantages there are several disadvantages to this model, including over-fitting. A remedy to this phenomenon is applying a boosting technique.

Boosting is one of the ensemble methods in Machine Learning. The very methodology was proposed as an answer to the question, weather two complex classes of learning problems were equivalent, strongly learnable and weakly learnable. The boosting technique allows weak models, which is better than a complete random guess, to be boosted in to an arbitrary strong model. AdaBoost is one among the most popular boosting methodologies widely used in Machine Learning. During the past few years, lots of alternatives and enhancements were proposed to various classes of boosting algorithms. In the current exercise Boosted Decision Tree Regression (BDTR) algorithm implemented in AzureML is used.

The BDTR implementation in AzureML uses an efficient implementation of Multiple Additive Regression Trees (MART) Gradient Boosting (GB). Gradient Boosting is a machine learning technique for regression problems. It builds each regression tree in a step-wise fashion, using a predefined loss function to measure the error in each step and correct for it in the next. Thus the prediction model is actually an ensemble of weaker prediction models. In regression problems, boosting builds a series of trees in in a step-wise fashion, and then selects the optimal tree using an arbitrary differentiable loss function [23].

### 3.2 Hyper Parameter Optimization

One of the key aspects of producing the best predictive model is selecting right value for the tuning parameters for the learning algorithms. Since there are no thumb rules in pre-determining the best parameter, it is advised to do a hyper parameter optimization by manual methods or the algorithmic way. In case of manual methods, *n* models with *m* parameter sets has to be developed, and keep track of the accuracy figures and then take a decision on which model has to be selected. At the same time there are algorithms and implementations available to handle this. A selected set of hyper parameters (either a small or large/exhaustive) can be supplied to an algorithm to fit *n* models with the *m* parameter combinations. Finally, the best set of parameters which yield the maximum performance is identified and entire training data is fitted/trained with the same parameters. The algorithm for Hyper Parameter Optimization is:

1. Define sets of model parameter values to evaluate
2. **for** each parameter set **do**
    **for** each resampling iteration **do**
        1. Hold-out specific samples
        2. [Optional] Pre-process the data
        3. Fit the model on the reminder
        4. Predict the hold-out samples
    **End**
    Calculate the average performance across hold-out predictions
3. End
4. Determine the optional parameter set
5. Fit the final model to all the training data using optional parameter set

The common methods for Hyper Parameter optimization are: Bayesian Optimization, Grid Search, Random Search and Gradient Boosting. A detailed discussion on the topic is available on paper by Bergstra et al [24][25]. In the current exercise the authors have used the Random Search technique to get the best model.

### 3.3 Modeling Electricity Consumption

The modeling exercise for the electricity consumption forecasting for smart meter data was done using Boosted Decision Tree Regression algorithm with Random Grid Search. The Random Grid Search was chosen to make sure that the best parameter applied to the model. A total of five iterations were done during the training phase. Mean Absolute Error was used to select the best model from the Grid Search results. The training and test data split was done using year wise separator. Data from 2010 to 2013 was used as the training set and data from year 2014 was kept as test set. As the current problem is a time-series forecasting, splitting the data by random 70-30 split or stratified split will corrupt the seasonal phenomena captured though various feature engineering techniques. The best model produced Normalized Root Mean Square Error (NRMSE) of 1.24. A detailed discussion on the Grid Search results and the final model is done in the results and discussion section.

A Decision Forest regression (DFR) [26] based model was also developed to create an ensemble model. The AzureML implementation of DFR is used for this. After obtaining the results from the BDTR and DFR a Linear Regression (LR) model was developed using the BDTR and DFR results as exploratory variable and the consumption as response. This technique is typically considered as model ensembling. These methods were proven to provide reliable results in many of the reported cases in Kaggle Data Science Challenges. Typically, this model ensembling will be done after selecting the various models after exhaustive or random grid based hyper parameter optimization exercises. The philosophy is that, multiple weak learner when unite can become a strong learner is adopted while modelling.

### 3.4 Model Evaluation

Root Mean Square Error (RMSE) and $R^2$ are selected as error metric. $R^2$ is calculated as the square of correlation between the observed *y* values and the predicted $\hat{y}$ values.

$$R^2 = \frac{\sum(\hat{y}i - \bar{y})^2}{\sum(yi - \bar{y})^2}$$

Otherwise it can be referred as the proportion of variation in the forecast variable that is accounted for by the regression model. If the predictions are close to the actual values, $R^2$ *should* to be close to 1. On the other hand, if the predictions are unrelated to the actual values, then $R^2 = 0$. In all cases, $R^2$ lies between 0 and 1. The use of RMSE is very common error metric for numerical predictions. Compared to the similar Mean Absolute Error, RMSE amplifies and severely punishes large errors.

$$RMSE = \sqrt{\frac{1}{n}(y_i - \hat{y}_i)^2}$$

For model selection RMSE was considered in the Random Grid Search and for test set evaluation $R^2$ was used.

## 4 RESULTS AND DISCUSSION
### 4.1 Results

The final data set used for modelling exercise consisted of following attributes: day hour index, is the day holiday, name of the day (such as Monday), whether the day is weekend or not, maximum consumption of the day, minimum consumption of the day, demand in hour minus one and demand in hour minus two. Three iterations of the training with BDTR has performed and evaluated the model with test data using the NRMSE score. The first iteration was done using five parameters combination to the BDTR and then 10, 15, 20 and finally 30 combinations. The combinations and results from the combinations in the training set is provided in the Appendix

A. The parameter combination 10, 15 and 20 model reported NRMSE of 1.20, the 30-parameter combination yielded NRMSE of 1.21 and 5 parameter combination yielded NRMSE of 1.24.

Table 2: NRMSE Scores in test data for models created with various parameter combinations

| Parameter Combinations | NRMSE in Test Data |
|---|---|
| 5 | 1.24 |
| 10 | 1.2 |
| 15 | 1.2 |
| 20 | 1.2 |
| 30 | 1.21 |

Following graph is prediction from the model for one-week period. Each point on horizontal axis is half an hour period and on vertical axis 1000 times scaled version of electrical consumption. The scaling has been done to model even minute details of the data.

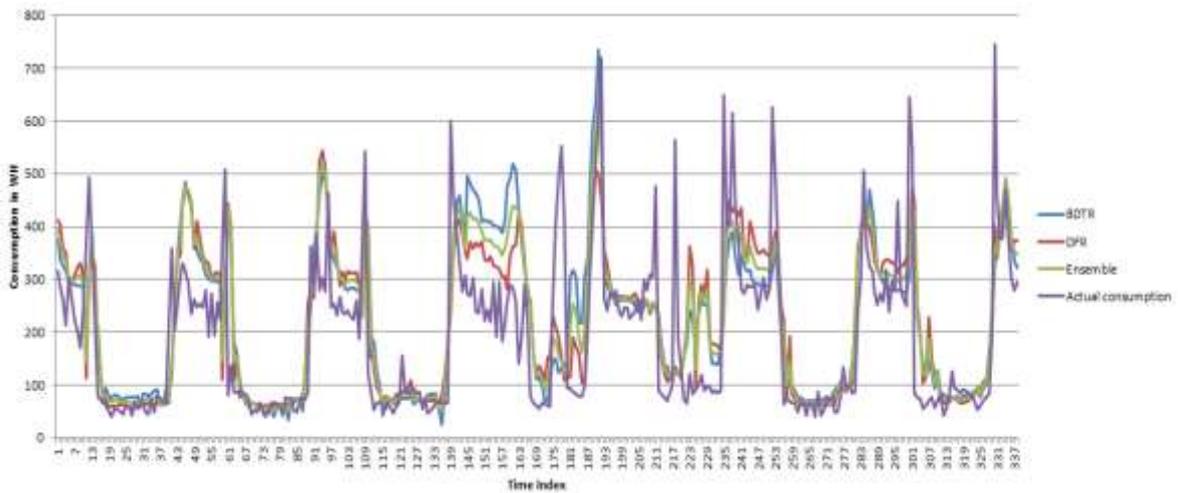

Figure: Sample plot for predicted vs. actual consumption.

A sample list of parameters used for training the BDTR is listed below. The actual larger list is provided in the Appendix B.

Table 3: Sample list of parameters used for training the BDTR

| **Number of Leaves** | **Minimum Leaf Instances** | **Learning Rate** | **Number of Trees** |
|---|---|---|---|
| 126 | 48 | 0.39997 | 482 |
| 124 | 48 | 0.391119 | 458 |
| 113 | 47 | 0.356604 | 443 |
| 111 | 45 | 0.336396 | 436 |
| 89 | 41 | 0.333128 | 396 |
| 86 | 38 | 0.300843 | 368 |
| 78 | 37 | 0.29927 | 362 |
| 67 | 32 | 0.291602 | 358 |

Initial result from a list of five parameters and the accuracy measures in training data is shown in the below table. A detailed list is available in the Appendix A.

Table 4: Parameter Tuning Results from 5 parameter combination

| Number of leaves | Minimum leaf instances | Learning rate | Number of trees | Mean Absolute Error | Root Mean Squared Error | Relative Absolute Error | Relative Squared Error | Coefficient of Determination |
|---|---|---|---|---|---|---|---|---|
| 2 | 4 | 0.064733 | 436 | 5.179087 | 8.945466 | 0.941751 | 0.735972 | 0.264028 |
| 19 | 1 | 0.042001 | 86 | 4.243727 | 7.442362 | 0.771668 | 0.509421 | 0.490579 |
| 6 | 47 | 0.148534 | 233 | 4.170603 | 7.281496 | 0.758371 | 0.487637 | 0.512363 |
| 36 | 7 | 0.333128 | 182 | 3.976567 | 6.989573 | 0.723088 | 0.449321 | 0.550679 |
| 59 | 27 | 0.391119 | 22 | 3.972663 | 7.005699 | 0.722378 | 0.451396 | 0.548604 |

## 4.2 Discussion

The exploratory analysis for this experiment reveled several interesting patterns in energy consumption. The data which was used was from an office building and the usage peaks were detected on working days and during the working hours. During the week-ends the consumption remained almost constant. The day shoulder effect was not much visible on office building data but for household consumption data from EMA Singapore, the pattern was just reverse. During the week-ends the consumption at day-time increases and spikes were visible in evening till late night. This observation helped in designing features such as hour -1 and hour -2 consumption features. Due to lack of adequate weather forecast data it has not been incorporated in the experiment. Some of ongoing research with similar data sets have included these features. Adding more features such as demographic information of household such as number of occupants with profession will help in producing reliable forecasting models. The weather data such as temperature, humidity, and rain will have an impact in the consumption patterns. The patterns observed were classified as climate event patterns, calendar event patterns and demography patterns. The demography patterns are the most interesting ones in the energy consumption data, such as a consumer with one or more youngsters/students shows peak consumption during the late-nights. At the same time, a household with homemakers and working professionals will have peak consumption during the early morning and late evening, with climatic seasonal patterns during the day and night time. Some of the climate event patterns will be varying from geography to geography and in sub-geographic regions too.

Following things can be concluded on electricity load forecasting from the research in this paper. Availability of supplementary metrological data and demographic information is crucial for producing reliable models. The analysis and treatment of categorical variables and the ability of selected algorithm and implementation to handle categorical data impacts the performance of the system. Hyper Parameter tuning with random and exhaustive grid option with granular parameter combinations can produce dependable model. At the same time this will depend upon the feature engineering, extraction and feature importance calculations. Even though the tree based ensemble models compute the feature importance internally, it is advised to perform a feature importance calculated with a base line model and build model with various feature importance techniques will be having a definite impact in the final model.

## 5 SUMMARY AND ACKNOWLEDGEMENTS

## 5.1 Summary

In this paper the authors assessed various challenges and techniques involved in electricity consumption forecasting. The feasibility of applying regression, tree and ensemble models were assessed along with various Data Science practices in data pre-processing and model tuning. The authors conducted an elaborate study on industry data and supplementary data required for the electricity forecasting and the challenges involved in feature engineering. An assessment on feature importance evaluation in time-series forecasting. The future of this experiment is to study the applications of Machine Learning in curtailment and demand response and personalization of usage plans for energy utility industry.

## 5.2  Acknowledgements

The authors gratefully acknowledge the motivation and support provided by Amit Singh, Sudipta Basu Ray and Sreeja Achuthan, EIM, Cognizant for providing their valuable suggestions and subject matter expertise in energy industry Machine Learning and Data Science. We would like to acknowledge the valuable suggestion provided by Cognizant associates during the course of this experiment. Any opinions, findings, and conclusions or recommendations expressed in this material are those of the author(s) and do not necessarily reflect the views of the Cognizant Technology Solution or its affiliates.

## 5.3  Competing interests

The authors declare that they have no competing interests. The authors declare that no proprietary information related to the authors, affiliated company or its approach, methodologies and IPR is discussed in this paper.

**Appendix A: Hyper Parameter Optimization Results with BDTR**

**Parameter Combination with 5 and NRMSE 1.24 in test data.**

| Number of leaves | Minimum leaf instances | Learning rate | Number of trees | Mean Absolute Error | Root Mean Squared Error | Relative Absolute Error | Relative Squared Error | Coefficient of Determination |
|---|---|---|---|---|---|---|---|---|
| 2 | 4 | 0.064733 | 436 | 5.179087 | 8.945466 | 0.941751 | 0.735972 | 0.264028 |
| 19 | 1 | 0.042001 | 86 | 4.243727 | 7.442362 | 0.771668 | 0.509421 | 0.490579 |
| 6 | 47 | 0.148534 | 233 | 4.170603 | 7.281496 | 0.758371 | 0.487637 | 0.512363 |
| 36 | 7 | 0.333128 | 182 | 3.976567 | 6.989573 | 0.723088 | 0.449321 | 0.550679 |
| 59 | 27 | 0.391119 | 22 | 3.972663 | 7.005699 | 0.722378 | 0.451396 | 0.548604 |

**Parameter Combination with 10 and NRMSE 1.20 in test data.**

| Number of leaves | Minimum leaf instances | Learning rate | Number of trees | Mean Absolute Error | Root Mean Squared Error | Relative Absolute Error | Relative Squared Error | Coefficient of Determination |
|---|---|---|---|---|---|---|---|---|
| 2 | 26 | 0.040069 | 159 | 5.27985 | 9.445324 | 0.960074 | 0.820519 | 0.179481 |
| 3 | 2 | 0.077932 | 202 | 4.660205 | 8.030981 | 0.847399 | 0.593188 | 0.406812 |
| 16 | 16 | 0.034768 | 48 | 4.510456 | 7.964228 | 0.820169 | 0.583368 | 0.416632 |
| 62 | 2 | 0.39997 | 94 | 4.033442 | 7.270072 | 0.73343 | 0.486108 | 0.513892 |
| 26 | 2 | 0.29927 | 358 | 3.990064 | 7.220218 | 0.725542 | 0.479464 | 0.520536 |
| 30 | 16 | 0.235847 | 27 | 3.99935 | 7.029924 | 0.727231 | 0.454524 | 0.545476 |
| 61 | 17 | 0.269985 | 178 | 4.007196 | 7.009512 | 0.728658 | 0.451888 | 0.548112 |
| 26 | 6 | 0.042918 | 337 | 3.919466 | 6.933737 | 0.712705 | 0.442171 | 0.557829 |
| 86 | 22 | 0.063668 | 87 | 3.844067 | 6.895167 | 0.698995 | 0.437265 | 0.562735 |
| 41 | 16 | 0.035407 | 368 | 3.84235 | 6.822146 | 0.698682 | 0.428053 | 0.571947 |
| | | | | | | | | |

**Parameter Combination with 15 and NRMSE 1.20 in test data.**

| Number of leaves | Minimum leaf instances | Learning rate | Number of trees | Mean Absolute Error | Root Mean Squared Error | Relative Absolute Error | Relative Squared Error | Coefficient of Determination |
|---|---|---|---|---|---|---|---|---|
| 2 | 26 | 0.040069 | 159 | 5.27985 | 9.445324 | 0.960074 | 0.820519 | 0.179481 |
| 4 | 6 | 0.111975 | 57 | 4.770994 | 8.264861 | 0.867545 | 0.628241 | 0.371759 |
| 33 | 1 | 0.037798 | 28 | 4.438059 | 7.918872 | 0.807005 | 0.576742 | 0.423258 |
| 5 | 37 | 0.030064 | 362 | 4.480355 | 7.818326 | 0.814696 | 0.56219 | 0.43781 |
| 60 | 2 | 0.041972 | 24 | 4.31793 | 7.726475 | 0.785161 | 0.549058 | 0.450942 |
| 17 | 13 | 0.062866 | 50 | 4.276761 | 7.509942 | 0.777675 | 0.518714 | 0.481286 |
| 25 | 41 | 0.097434 | 32 | 4.191364 | 7.432731 | 0.762146 | 0.508103 | 0.491897 |
| 62 | 2 | 0.39997 | 94 | 4.033442 | 7.270072 | 0.73343 | 0.486108 | 0.513892 |
| 26 | 2 | 0.29927 | 358 | 3.990064 | 7.220218 | 0.725542 | 0.479464 | 0.520536 |
| 30 | 16 | 0.235847 | 27 | 3.99935 | 7.029924 | 0.727231 | 0.454524 | 0.545476 |
| 32 | 6 | 0.252099 | 270 | 3.931318 | 6.989473 | 0.71486 | 0.449308 | 0.550692 |
| 33 | 9 | 0.046704 | 146 | 3.952085 | 6.97901 | 0.718636 | 0.447964 | 0.552036 |
| 54 | 19 | 0.336396 | 51 | 3.918383 | 6.916063 | 0.712508 | 0.439919 | 0.560081 |
| 26 | 15 | 0.143552 | 396 | 3.883652 | 6.886089 | 0.706193 | 0.436114 | 0.563886 |

| Number of leaves | Minimum leaf instances | Learning rate | Number of trees | Mean Absolute Error | Root Mean Squared Error | Relative Absolute Error | Relative Squared Error | Coefficient of Determination |
|---|---|---|---|---|---|---|---|---|
| 41 | 16 | 0.035407 | 368 | 3.84235 | 6.822146 | 0.698682 | 0.428053 | 0.571947 |

**Parameter Combination with 20 and NRMSE 1.20 in test data.**

| Number of leaves | Minimum leaf instances | Learning rate | Number of trees | Mean Absolute Error | Root Mean Squared Error | Relative Absolute Error | Relative Squared Error | Coefficient of Determination |
|---|---|---|---|---|---|---|---|---|
| 2 | 4 | 0.264257 | 24 | 5.281863 | 9.409552 | 0.96044 | 0.814316 | 0.185684 |
| 4 | 2 | 0.135326 | 44 | 4.791843 | 8.297317 | 0.871336 | 0.633185 | 0.366815 |
| 4 | 6 | 0.111975 | 57 | 4.770994 | 8.264861 | 0.867545 | 0.628241 | 0.371759 |
| 33 | 1 | 0.037798 | 28 | 4.438059 | 7.918872 | 0.807005 | 0.576742 | 0.423258 |
| 5 | 37 | 0.030064 | 362 | 4.480355 | 7.818326 | 0.814696 | 0.56219 | 0.43781 |
| 60 | 2 | 0.041972 | 24 | 4.31793 | 7.726475 | 0.785161 | 0.549058 | 0.450942 |
| 10 | 7 | 0.054386 | 102 | 4.35105 | 7.59393 | 0.791183 | 0.530381 | 0.469619 |
| 10 | 38 | 0.035918 | 190 | 4.307129 | 7.564457 | 0.783197 | 0.526272 | 0.473728 |
| 17 | 13 | 0.062866 | 50 | 4.276761 | 7.509942 | 0.777675 | 0.518714 | 0.481286 |
| 25 | 41 | 0.097434 | 32 | 4.191364 | 7.432731 | 0.762146 | 0.508103 | 0.491897 |
| 18 | 1 | 0.05442 | 98 | 4.173567 | 7.323335 | 0.75891 | 0.493257 | 0.506743 |
| 126 | 9 | 0.029807 | 46 | 3.970658 | 7.182756 | 0.722014 | 0.474501 | 0.525499 |
| 32 | 6 | 0.252099 | 270 | 3.931318 | 6.989473 | 0.71486 | 0.449308 | 0.550692 |
| 25 | 25 | 0.167076 | 125 | 3.951615 | 6.985111 | 0.718551 | 0.448747 | 0.551253 |
| 33 | 9 | 0.046704 | 146 | 3.952085 | 6.97901 | 0.718636 | 0.447964 | 0.552036 |
| 6 | 15 | 0.356604 | 482 | 3.950589 | 6.922986 | 0.718364 | 0.4408 | 0.5592 |
| 54 | 19 | 0.336396 | 51 | 3.918383 | 6.916063 | 0.712508 | 0.439919 | 0.560081 |

| Number of leaves | Minimum leaf instances | Learning rate | Number of trees | Mean Absolute Error | Root Mean Squared Error | Relative Absolute Error | Relative Squared Error | Coefficient of Determination |
|---|---|---|---|---|---|---|---|---|
| 26 | 15 | 0.143552 | 396 | 3.883652 | 6.886089 | 0.706193 | 0.436114 | 0.563886 |
| 66 | 9 | 0.149896 | 105 | 3.810655 | 6.820037 | 0.692919 | 0.427788 | 0.572212 |
| 67 | 11 | 0.03337 | 218 | 3.801592 | 6.81379 | 0.691271 | 0.427005 | 0.572995 |

**Parameter Combination with 30 and NRMSE 1.21 in test data.**

| Number of leaves | Minimum leaf instances | Learning rate | Number of trees | Mean Absolute Error | Root Mean Squared Error | Relative Absolute Error | Relative Squared Error | Coefficient of Determination |
|---|---|---|---|---|---|---|---|---|
| 2 | 4 | 0.264257 | 24 | 5.281863 | 9.409552 | 0.96044 | 0.814316 | 0.185684 |
| 2 | 4 | 0.240191 | 38 | 5.257497 | 9.275193 | 0.956009 | 0.791227 | 0.208773 |
| 13 | 6 | 0.045541 | 25 | 4.734457 | 8.419569 | 0.860901 | 0.651981 | 0.348019 |
| 4 | 2 | 0.135326 | 44 | 4.791843 | 8.297317 | 0.871336 | 0.633185 | 0.366815 |
| 3 | 2 | 0.099039 | 129 | 4.728256 | 8.163223 | 0.859773 | 0.612884 | 0.387116 |
| 5 | 1 | 0.08631 | 85 | 4.624512 | 8.001211 | 0.840909 | 0.588798 | 0.411202 |
| 3 | 10 | 0.117867 | 165 | 4.583054 | 7.897454 | 0.83337 | 0.573627 | 0.426373 |
| 8 | 48 | 0.027429 | 185 | 4.488463 | 7.87454 | 0.81617 | 0.570303 | 0.429697 |
| 32 | 48 | 0.057697 | 24 | 4.350257 | 7.800561 | 0.791039 | 0.559638 | 0.440362 |
| 5 | 1 | 0.300843 | 35 | 4.510288 | 7.760104 | 0.820139 | 0.553848 | 0.446152 |
| 5 | 3 | 0.048922 | 249 | 4.432146 | 7.658441 | 0.805929 | 0.539431 | 0.460569 |
| 10 | 7 | 0.054386 | 102 | 4.35105 | 7.59393 | 0.791183 | 0.530381 | 0.469619 |
| 10 | 38 | 0.035918 | 190 | 4.307129 | 7.564457 | 0.783197 | 0.526272 | 0.473728 |
| 15 | 12 | 0.060661 | 65 | 4.265019 | 7.492609 | 0.775539 | 0.516323 | 0.483677 |
| 8 | 1 | 0.03274 | 337 | 4.274652 | 7.440833 | 0.777291 | 0.509212 | 0.490788 |

| 6 | 6 | 0.043036 | 348 | 4.263654 | 7.403377 | 0.775291 | 0.504098 | 0.495902 |
| 18 | 1 | 0.05442 | 98 | 4.173567 | 7.323335 | 0.75891 | 0.493257 | 0.506743 |
| 126 | 9 | 0.029807 | 46 | 3.970658 | 7.182756 | 0.722014 | 0.474501 | 0.525499 |
| 38 | 32 | 0.152104 | 45 | 3.982146 | 7.074468 | 0.724103 | 0.460302 | 0.539698 |
| 124 | 1 | 0.153058 | 48 | 3.823641 | 7.057243 | 0.695281 | 0.458063 | 0.541937 |
| 89 | 1 | 0.102658 | 443 | 3.853832 | 7.047303 | 0.70077 | 0.456774 | 0.543226 |
| 113 | 1 | 0.103661 | 33 | 3.850595 | 7.032107 | 0.700182 | 0.454806 | 0.545194 |
| 11 | 45 | 0.291602 | 458 | 4.032023 | 7.028673 | 0.733172 | 0.454362 | 0.545638 |
| 25 | 25 | 0.167076 | 125 | 3.951615 | 6.985111 | 0.718551 | 0.448747 | 0.551253 |
| 6 | 15 | 0.356604 | 482 | 3.950589 | 6.922986 | 0.718364 | 0.4408 | 0.5592 |
| 78 | 8 | 0.220409 | 104 | 3.879945 | 6.879155 | 0.705519 | 0.435237 | 0.564763 |
| 111 | 9 | 0.212698 | 56 | 3.819187 | 6.830715 | 0.694471 | 0.429129 | 0.570871 |
| 66 | 9 | 0.149896 | 105 | 3.810655 | 6.820037 | 0.692919 | 0.427788 | 0.572212 |
| 67 | 11 | 0.03337 | 218 | 3.801592 | 6.81379 | 0.691271 | 0.427005 | 0.572995 |
| 58 | 12 | 0.123249 | 118 | 3.797312 | 6.789049 | 0.690493 | 0.423909 | 0.576091 |

**Appendix B: List of Parameters Tried for Training the Boosted Decision Tree**

| Number of Leaves | Minimum Leaf Instances | Learning Rate | Number of Trees |
|---|---|---|---|
| 126 | 48 | 0.39997 | 482 |
| 124 | 48 | 0.391119 | 458 |
| 113 | 47 | 0.356604 | 443 |
| 111 | 45 | 0.336396 | 436 |
| 89 | 41 | 0.333128 | 396 |
| 86 | 38 | 0.300843 | 368 |
| 78 | 37 | 0.29927 | 362 |
| 67 | 32 | 0.291602 | 358 |
| 66 | 27 | 0.269985 | 348 |
| 62 | 26 | 0.264257 | 337 |
| 61 | 25 | 0.252099 | 337 |
| 60 | 22 | 0.240191 | 270 |
| 59 | 19 | 0.235847 | 249 |
| 58 | 17 | 0.220409 | 233 |
| 54 | 16 | 0.212698 | 218 |
| 41 | 16 | 0.167076 | 202 |
| 38 | 16 | 0.153058 | 190 |
| 36 | 15 | 0.152104 | 185 |
| 33 | 15 | 0.149896 | 182 |
| 33 | 13 | 0.148534 | 178 |
| 32 | 12 | 0.143552 | 165 |
| 32 | 12 | 0.135326 | 159 |
| 30 | 11 | 0.123249 | 146 |
| 26 | 10 | 0.117867 | 129 |
| 26 | 9 | 0.111975 | 125 |
| 26 | 9 | 0.103661 | 118 |
| 25 | 9 | 0.102658 | 105 |
| 25 | 9 | 0.099039 | 104 |
| 19 | 8 | 0.097434 | 102 |
| 18 | 7 | 0.08631 | 98 |
| 17 | 7 | 0.077932 | 94 |
| 16 | 6 | 0.064733 | 87 |
| 15 | 6 | 0.063668 | 86 |

| 13 | 6 | 0.062866 | 85 |
| --- | --- | --- | --- |
| 11 | 6 | 0.060661 | 65 |
| 10 | 6 | 0.057697 | 57 |
| 10 | 4 | 0.05442 | 56 |
| 8 | 4 | 0.054386 | 51 |
| 8 | 4 | 0.048922 | 50 |
| 6 | 3 | 0.046704 | 48 |
| 6 | 2 | 0.045541 | 48 |
| 6 | 2 | 0.043036 | 46 |
| 5 | 2 | 0.042918 | 45 |
| 5 | 2 | 0.042001 | 44 |
| 5 | 2 | 0.041972 | 38 |
| 5 | 2 | 0.040069 | 35 |
| 4 | 1 | 0.037798 | 33 |
| 4 | 1 | 0.035918 | 32 |
| 3 | 1 | 0.035407 | 28 |
| 3 | 1 | 0.034768 | 27 |
| 3 | 1 | 0.03337 | 25 |
| 2 | 1 | 0.03274 | 24 |
| 2 | 1 | 0.030064 | 24 |
| 2 | 1 | 0.029807 | 24 |
| 2 | 1 | 0.027429 | 22 |